\documentclass[letter,rmp,twocolumn,amsmath,amssymb]{revtex4}
\usepackage{graphicx}
\usepackage{dcolumn}
\usepackage{bm}

\begin{document}

\newcommand{\be}{\begin{equation}}
\newcommand{\ee}{\end{equation}}
\newcommand{\bes}{\begin{equation}}
\newcommand{\ees}{\nonumber\end{equation}}
\newcommand{\bea}{\begin{eqnarray}}
\newcommand{\eea}{\end{eqnarray}}
\newcommand{\uv}[1]{\mathbf{\hat{#1}}}
\newcommand{\curl}[1]
{\mathbf{\nabla}\times\mathbf{#1}}
\newcommand{\dvg}[1]
{\mathbf{\nabla}\cdot\mathbf{#1}}
\newcommand{\pd}[2]
{{{\partial #1} \over {\partial #2}}}
\newcommand{\pdt}[2]
{{{\partial^{2} #1} \over {\partial #2 ^{2}}}}

\newcommand{\m}[1]
{\mathbf{#1}}

\preprint{Draft}

\title{Dual-frequency ferromagnetic resonance}

\author{Y. Guan\email{yg2111@columbia.edu} and W. E. Bailey}
\affiliation{Materials Science Program, Department of Applied
Physics, Columbia University, New York, New York 10027}

\date{\today}

\begin{abstract}

We describe a new experimental technique to investigate coupling
effects between different layers or modes in ferromagnetic
resonance (FMR). Dual FMR frequencies are excited (2-8 GHz)
simultaneously and detected selectively in a broadband RF circuit,
using lock-in amplifier detection at separate modulation
frequencies.

\end{abstract}

\maketitle

\section{Introduction}

Ferromagnetic resonance (FMR) precession of ferromagnetic alloys
and heterostructures is technologically important since it
determines the GHz dynamics of spin electronic devices. Recently,
novel long-ranged dynamic coupling mechanisms between layers in
heterostructures, through the transient excitation of spin
currents, have been observed experimentally\cite{Tserkovnyak-rmp}.
Some of these processes could be elucidated if motions of
individual layers can be excited independently, allowing the
effects of higher amplitude motion at one layer to be
characterized in the resonance at an opposite layer.

Experiments on driven FMR modes in ferromagnetic multilayers so
far have typically focused on the situation with single drive
excitation. Dual-drive excitations cannot be found in the
literature except in a few studies of nonlinear interactions
between magnetostatic waves in yttrium-iron-garnet
films\cite{mar-jap1, mar-jap2}. We have developed a dual-frequency
FMR technique to study interactions between FMR modes. Through the
adjustment of pumping power at mode 1 (excited at $f_{1}$), we can
examine effects on FMR line intensity, width, or shape at mode 2
(excited at $f_{2}$).

In this article, we describe the apparatus for dual-frequency FMR
measurements. FMR is excited at two frequencies using dual RF
sources, combined at a power divider, and delivered to the
ferromagnetic sample using a broadband coplanar waveguide (CPW).
The response of individual FMR modes is detected by modulating
each RF source at a separate low modulation frequency and locking
in to these two frequencies with separate lock-in amplifiers. The
technique is validated with a Ni$_{81}$Fe$_{19}$ thin film sample,
where dual-frequency FMR is proved to be selective of resonance
frequency. With this technique, investigations have been made in a
Ni$_{81}$Fe$_{19}$/Cu/Co$_{93}$Zr$_{7}$ multilayer, where some
influence of pumped Ni$_{81}$Fe$_{19}$ precession on the
Co$_{93}$Zr$_{7}$ resonance may be seen.

\section{Apparatus}

The block diagram of the experimental setup of dual-frequency FMR
is shown in Fig. 1, where broadband (2-8 GHz) swept-field
ferromagnetic resonance (FMR) is measured. Lock-in amplifier
detection through frequency modulation is used in our FMR
measurements. Apart from the dual frequency capability, our
measurements are similar to conventional FMR cavity
measurements\cite{heinrich}, although network analyzers are often
used instead of lock-in amplifier for phase-sensitive detection in
conventional broadband FMR spectrometers\cite{Denysenkov-rsi}.

The microwave sources consist of two independently controllable
frequency sweepers: RF source 1 and RF source 2. RF source 1 is a
home-built, fixed-frequency source at 2.3 GHz ($f_{1}$), variable
in output from -50 to 22 dBm. We used a filtered amplified higher
harmonic of a 88 MHz signal from a Waveteck 3000 signal with FM
capability. RF source 2 is a Wiltron 6668B sweep generator
operated in cw mode, with tunable frequency ($f_{2}$) in the range
of 0-40 GHz, variable output from -50 to 15 dBm, and FM
capability. The dual microwave generators are used to generate
dual rf signals ($f_{1}$, $f_{2}$) simultaneously. The two
independent rf sources modulated at separate frequencies
($f^{1}_{mod}$, $f^{2}_{mod}$) are combined with a Anaren 42020
broadband power combiner/divider (2-8 GHz), which attenuates each
input by 3 dB in transmission, and then delivered to the
ferromagnetic thin film sample through a lithographic coplanar
waveguide (CPW) with a 100 $\mu$m center conductor width. The RF
field delivery configuration here is similar to that used in PIMM
(pulsed inductive microwave magnetometer)
measurements\cite{kos-rsi}, substituting cw microwave sources for
the pulse generator.

The thin film sample is placed film-side down onto the top of the
CPW, with a thin layer of photoresist spin-coated onto the film to
prevent it from shorting the CPW. The CPW, used to pump the thin
film sample with combined microwave (2-8 GHz) excitations from two
synthesized microwave generators, is constructed by standard
lithographic techniques and placed inside a Fe core electromagnet.
The electromagnet has a gap of $\sim$ 2 cm and a maximum field of
$\sim$ 1.0 Tesla at 40 Ampere, controlled by parallel
KEPCO-BOP-20/20M current sources. The Fe core electromagnet
provides applied magnetic bias field along the center conductor of
CPW, which is measured directly using a transverse hall probe
monitored by a Lakeshore 421 Gaussmeter.

Transmitted rf signals through the magnetic thin film sample are
detected at a microwave diode (0-18 GHz), the output of which is
sent to the $A$ inputs of two lock-in amplifiers (lock-in 1 of
SR810, and lock-in 2 of SR830). The $sine$ outputs of the lock-in
amplifiers provide the modulation to the transmitted rf signals at
separate modulation frequencies: $f^{1}_{mod}$ ($\sim$ 120 Hz) for
$f_{1}$, and $f^{2}_{mod}$ ($\sim$ 540 Hz) for $f_{2}$. A
microcomputer equipped with a GPIB bus communicates with the
SR810/830 and the Wiltron 6668B; analog inputs ($\pm$10 V) at the
SR830 read the corrected output of the Lakeshore 421 Gaussmeter,
and parallel analog outputs ($\pm$5 V) control the Kepco power
supplies. All these features allow the system to be fully
automated.

In operation, the dual-frequency FMR measurement can sweep
frequencies at $f_{2}$ and fields $H_{B}$, while monitoring the
diode signals at $f^{1}_{mod}$ ($f_{1}$) and $f^{2}_{mod}$
($f_{2}$). Power levels at $f_{1}$ are set manually through a
variable attenuator. A high-precision (Keithley 2000) DMM, also
under GPIB control, was available to monitor the DC diode signals
in the same measurements.

\begin{figure}[h]
\includegraphics[width=\columnwidth]{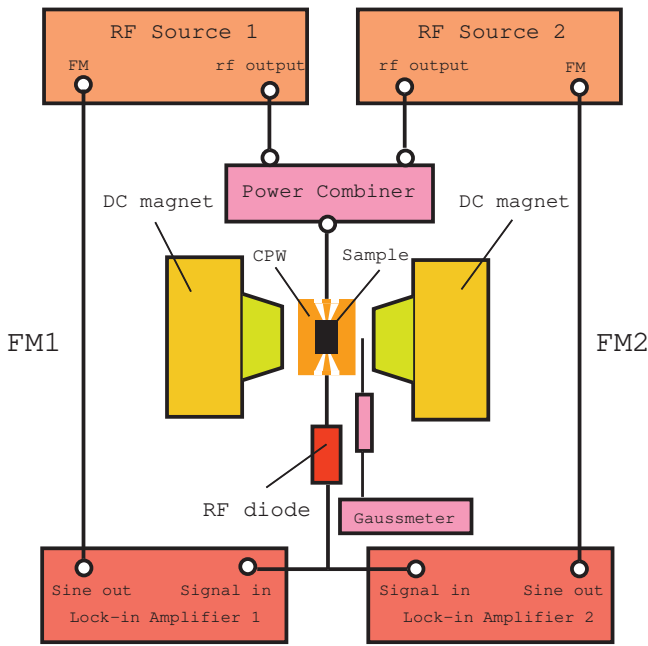}
\caption{(Color online) Block diagram of the experimental setup of
dual-frequency FMR.}
\end{figure}

\section{Experiment}

A single layer sample of Ni$_{81}$Fe$_{19}$(25 nm) and a trilayer
structure of Ni$_{81}$Fe$_{19}$(25 nm)/Cu(20
nm)/Co$_{93}$Zr$_{7}$(25 nm), were grown by UHV magnetron
sputtering from alloy targets at a base pressure of
4$\times$10$^{-9}$ torr onto Si/SiO$_{2}$ substrates. A 20 Oe
deposition field was applied to induce unidirectional anisotropy
in the film plane. More details on deposition conditions can be
found in our previous work\cite{guan-jap}.

Conventional broadband FMR measurements (RF source 1 is off, and
RF source 2 is on) were made on both Ni$_{81}$Fe$_{19}$ and
Ni$_{81}$Fe$_{19}$/Cu/Co$_{93}$Zr$_{7}$ thin film samples at
several selected frequencies (from 0 to 5 GHz) to determine the
Kittel relations\cite{kittel} between microwave frequency
(${\omega_{p}}/{{2\pi}}$) and resonance field ($H_{res}$). For
in-plane magnetization in a thin film, the Kittel relation can be
expressed as\cite{kittel}:
\begin{equation}
\omega_{p}^2 \approx \mu_{0}^{2}\gamma^{2}M_{s} (H_{res}+H_{k}),
\end{equation}
where $H_{k}$ is an effective field due to anisotropy.

The Ni$_{81}$Fe$_{19}$ thin film sample was then measured using
dual-frequency FMR, where two different FMR modes were excited
independently, using separate resonance frequencies ($f_{1}$ = 2.3
GHz, $f_{2}$ = 4.5 GHz) at fixed field. In the
Ni$_{81}$Fe$_{19}$/Cu/Co$_{93}$Zr$_{7}$ trilayer sample, both the
Ni$_{81}$Fe$_{19}$ layer and the Co$_{93}$Zr$_{7}$ layer were
excited near resonance simultaneously, choosing $H_{res}$ for
$f_{1}$ = 2.3 GHz, through the adjustment of $f_{2}$ (to 3.8 GHz).
The effects of variable power at $f_{1}$ on the Co$_{93}$Zr$_{7}$
resonance at $f_{2}$ were investigated.

\section{Results and discussion}

The results of conventional broadband FMR measurements on both
Ni$_{81}$Fe$_{19}$ and Ni$_{81}$Fe$_{19}$/Cu/Co$_{93}$Zr$_{7}$
thin film samples are presented in Fig. 2. We show
$(\omega_{p}/{2\pi})^2$ as a function of $H_{res}$. $f_{1}$ and
$f_{2}$ denote the selected frequencies of excited FMR modes in
dual-frequency FMR measurements on the same samples (shown in Fig.
3 and Fig. 4). Solid lines are linear fits based on Eq. (1). Two
branches were observed for the
Ni$_{81}$Fe$_{19}$/Cu/Co$_{93}$Zr$_{7}$ thin film sample,
corresponding mostly to the separate resonances of the
Ni$_{81}$Fe$_{19}$ layer and the Co$_{93}$Zr$_{7}$ layer, with
some weak ferromagnetic coupling ($\sim$ 5 Oe).

\begin{figure}[h]
\includegraphics[width=\columnwidth]{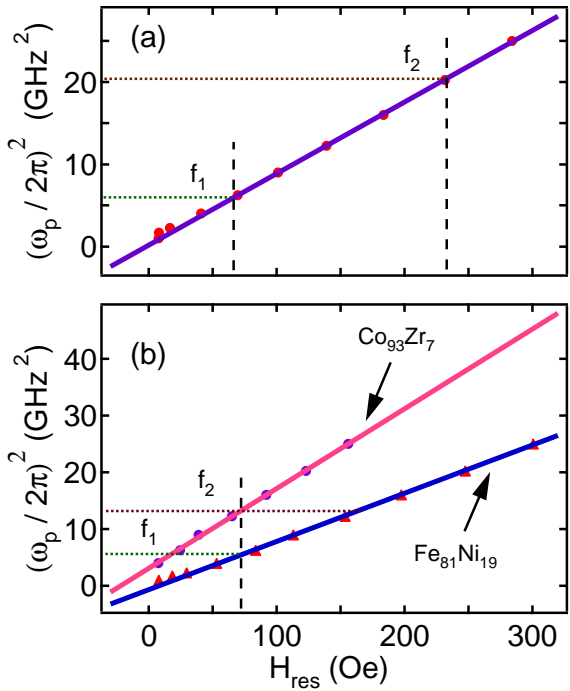}
\caption{(Color online) (a) Kittel plot of the single layer of
Ni$_{81}$Fe$_{19}$ thin film. (b) Kittel plot of the trilayer
structure of Ni$_{81}$Fe$_{19}$/Cu/Co$_{93}$Zr$_{7}$ thin film.
Solid lines are linear fits.}
\end{figure}

Validation of frequency selectivity in dual-frequency FMR is shown
in Fig. 3. Two different FMR modes were excited in the
Ni$_{81}$Fe$_{19}$ thin film sample at two selective resonance
frequencies ($f_{1}$ = 2.3 GHz, $f_{2}$ = 4.5 GHz) and then
detected selectively using dual lock-in amplifiers (lock-in 1,
lock-in 2), respectively. Two absorption peaks were observed for
the total transmitted signal, where the positions and widths of
the two absorption peaks correspond closely with those found
selectively at $f_{1}$ and $f_{2}$.

\begin{figure}[h]
\includegraphics[width=\columnwidth]{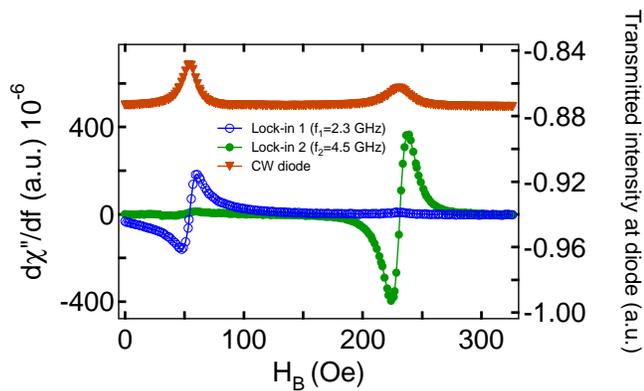}
\caption{(Color online) Dual-frequency FMR spectra ($f_{1}$ = 2.3
GHz, $f_{2}$ = 4.5 GHz) of Ni$_{81}$Fe$_{19}$ thin film using dual
lock-in detection.}
\end{figure}

Dual-frequency FMR measurement on the
Ni$_{81}$Fe$_{19}$/Cu/Co$_{93}$Zr$_{7}$ trilayer thin film sample
is presented in Fig. 4, where the effects of pumped
Ni$_{81}$Fe$_{19}$ precession at $f_{1}$ = 2.3 GHz were
investigated on the Co$_{93}$Zr$_{7}$ resonance at $f_{2}$ = 3.8
GHz. This frequency (f1) sets the Ni$_{81}$Fe$_{19}$ layer into
FMR at $H_{B}$ = 72.5 Oe. With rf power of RF source 2 fixed at a
low value (-5 dBm), we varied the pumping power of RF source 1,
from -20 dBm to +10 dBm, measured at the diode. As shown in Fig.
4, two separate FMR modes were observed for each value of the
pumping power of RF source 1, where the mode at the low-field side
corresponded primarily to the resonance of the Co$_{93}$Zr$_{7}$
layer. Some variations can be seen in both the line intensity and
the symmetry of the Co$_{93}$Zr$_{7}$ FMR modes as a function of
power pumped into the Ni$_{81}$Fe$_{19}$ resonance. It is also
evident that the high power excited at $f_{1}$ has no discernible
influence on the Ni$_{81}$Fe$_{19}$ resonance measured at $f_{2}$,
as expected. However, a quantitative estimate of any dynamic
coupling between Ni$_{81}$Fe$_{19}$ and Co$_{93}$Zr$_{7}$ layers
could not be made, as the residual influence of the static ($\sim$
5 Oe) coupling could not be excluded. It is additionally important
to drive the Ni$_{81}$Fe$_{19}$ resonance symmetrically over the
low-field region, which requires a second variable frequency
source at $f_{1}$.

\begin{figure}[h]
\includegraphics[width=\columnwidth]{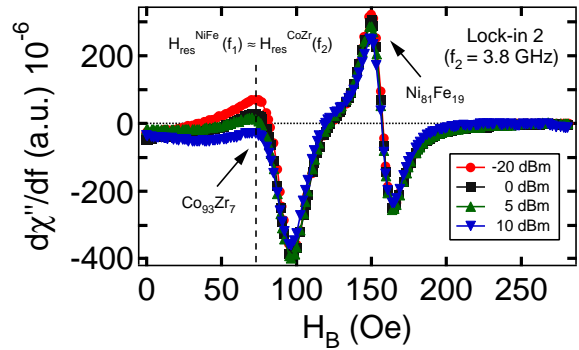}
\caption{(Color online) Dual-frequency FMR spectra at $f_{2}$ =
3.8 GHz as a function of pumped power at $f_{1}$ = 2.3 GHz in
Ni$_{81}$Fe$_{19}$(25 nm)/Cu(20 nm)/Co$_{93}$Zr$_{7}$(25 nm) thin
film. The field $H_{B}$ for simultaneous FMR of Ni$_{81}$Fe$_{19}$
and Co$_{93}$Zr$_{7}$ is indicated.}
\end{figure}

By using dual-frequency FMR technique, we can selectively excite
several different FMR modes in a ferromagnetic alloy and
heterostructures, and the nature of coupling interactions between
multiple FMR modes could thus be probed.We remark that the
technique presented could be extended to fixed-field,
swept-frequency measurement, or to track the Kittel resonance at
$f_{1}$ as field $H_{res}$ is swept. With much higher power
amplification and some cooling capability, nonlinear interactions
between magnetostatic spin-wave modes (MSSW) could be studied as
well. Finally, the frequency range can be extended to 20 GHz, or
even to 40 GHz, through the use of different power dividers.

\section{Conclusion}

A new experimental technique, dual-frequency ferromagnetic
resonance (FMR), has been developed to investigate coupling
effects between different FMR modes. This new technique is able to
excite different FMR modes simultaneously and independently, while
separating the properties of each.

\section{Acknowledgements}
This work was partially supported by the Army Research Office with
Grant No. ARO-43986-MS-YIP, and the National Science Foundation
with Grant No. NSF-DMR-02-39724. This work has used the shared
experimental facilities that are supported primarily by the
MRSEC(Columbia) program of the National Science Foundation under
Contract No. NSF-DMR-0213574.

\end{document}